\documentclass[aps,showpacs,prl,toolkits,twocolumn]{revtex4}
\usepackage{epsfig,amssymb,amsfonts,amsmath,graphicx}

\makeatletter
\renewcommand{\@makefntext}[1]{\parindent=1em\noindent\hbox to 1.8em
{\hss$^{\@thefnmark}$}#1}
\renewcommand{\@footnotemark}{\hbox{\mathsurround=0pt$^{\@thefnmark}$}}

\makeatother

\begin{document}
\title{Chirally symmetric and confining dense  matter with a diffused
quark Fermi surface.}
\author{ L. Ya. Glozman, V. K. Sazonov and R. F. Wagenbrunn}
\affiliation{Institute for
 Physics, Theoretical Physics branch, University of Graz, Universit\"atsplatz 5,
A-8010 Graz, Austria}

\newcommand{\be}{\begin{equation}}
\newcommand{\bea}{\begin{eqnarray}}
\newcommand{\ee}{\end{equation}}
\newcommand{\eea}{\end{eqnarray}}
\newcommand{\ds}{\displaystyle}
\newcommand{\low}[1]{\raisebox{-1mm}{$#1$}}
\newcommand{\loww}[1]{\raisebox{-1.5mm}{$#1$}}
\newcommand{\lmn}{\mathop{\sim}\limits_{n\gg 1}}
\newcommand{\vpint}{\int\makebox[0mm][r]{\bf --\hspace*{0.13cm}}}
\newcommand{\too}{\mathop{\to}\limits_{N_C\to\infty}}
\newcommand{\vp}{\varphi}
\newcommand{\vx}{{\vec x}}
\newcommand{\vy}{{\vec y}}
\newcommand{\vz}{{\vec z}}
\newcommand{\vk}{{\vec k}}
\newcommand{\vq}{{\vec q}}
\newcommand{\vpp}{{\vec p}}
\newcommand{\vn}{{\vec n}}
\newcommand{\vg}{{\vec \gamma}}

%\Large
\begin{abstract}

It is possible that at low temperatures and  large density there
exists a confining matter with restored chiral symmetry, just after
the dense nuclear matter with broken chiral symmetry.
Such a phase has  sofar been studied within a 
confining and chirally symmetric model assuming a rigid quark Fermi surface.
In the confining quarkyonic matter, however, near the Fermi surface quarks
group into color-singlet baryons. Interaction between quarks
 leads to a diffusion of the quark Fermi surface. Here
we study effects of such diffusion and verify that it does not
destroy a possible existence of a confining but chirally symmetric
matter at low temperatures.
\end{abstract}

\maketitle

\section{Introduction}

The QCD phase diagram is an old and very intriguing question.
What will happen with the strongly interacting matter at
large temperatures and/or densities? The most interesting question
is the interconnection of the deconfinement and the chiral restoration
phase transitions (crossovers). For many years it was considered as 
almost selfevident that the deconfinement and chiral restoration
lines on the QCD phase diagram coincide. The reason for such
expectation was the belief that in the confining mode mass of hadrons is
directly related to the quark condensate of the vacuum. Consequently,
beyond the chiral restoration line hadrons cannot exist and the QCD
matter should be in a deconfined plasma form. At low densities it is
indeed established on the lattice that both chiral restoration
and deconfinement crossovers coincide or
are rather close to each other \cite{fodor,
karsch}. What happens with these lines at larger densities is unknown.

In the large $N_c$ world at low temperatures confinement persists
up to arbitrary large densities \cite{pisarski}. This is because
both the quark - antiquark and quark - quark hole loops are
suppressed at large $N_c$. Consequently, there is no back reaction
of quarks on gluonic dynamics (no   Debye screening of the confining
gluonic field) and confinement persists like in vacuum. In such
system the only allowed excitation modes are of the color-singlet
hadronic type. The uncorrelated single quark excitations are not
allowed. In this case it is possible to define a quarkyonic matter
as a dense confining matter with baryonic excitation modes \cite{pisarski}.

In the real $N_c=3$ world at some large critical density the confining
gluonic field might be screened and a deconfining transition (crossover)
would appear. Then a key question is how big is this critical density?
Lattice simulations for the $N_c=2$ QCD suggest that at low temperatures
the deconfinement transition happens at densities  $\sim 100$ times
bigger than the normal nuclear matter density \cite{Hands}. Since the $N_c=3$
world is between the two known limiting cases ($N_c=2; N_c=\infty$),
we expect that at $N_c=3$ the deconfinement at low temperatures
happens at the very high densities, much larger than can be achieved
in our laboratories or in the neuteron stars.

Note,  by definition the quarkyonic matter is a dense cold matter
with confinement. Nothing can be apriori concluded about existence or
nonexistence of the chiral restoration phase transition within such
a dense matter. 
If the chiral restoration transition does exist within the
quarkyonic matter, then it would imply that
at some conditions there exists a QCD matter with confinement and 
with unbroken chiral symmetry (we imply for simplicity the chiral limit).
The mass  origin in such a matter is obviously not related
to dynamical chiral symmetry breaking in the vacuum.

As mentioned above, such a possibility  had not even 
been considered in the past on apriori grounds. 
Indeed, the old Casher argument \cite{Casher} claims that
the chiral symmetry breaking is required for quarks to
be confined.
In addition, it is  understood
that chiral symmetry breaking in the vacuum is
 important for the mass generation of hadrons such as 
$N$, $\rho$, or $\pi$. Then, naively, hadrons with nonzero mass cannot
exist in a world with unbroken chiral symmetry. 

However, the Casher argument is not general and 
can be easily bypassed \cite{G0}.
Recent lattice simulations have convincingly demonstrated \cite{LS}
that in the world without the low-lying eigenmodes of the Dirac
operator (i.e. with the artificially restored chiral symmetry) 
hadrons still exist and confinement persists.  
Finally, if effective
chiral restoration in highly excited hadrons  is correct
\cite{G1,G2,G3,G4},
then it is possible to have hadrons with the mass that is not directly related 
to the quark condensate of the vacuum.

A key question is to clarify whether
existence of confining but chirally symmetric dense, cold matter
is possible. We cannot solve QCD and answer this question from
first principles. In this situation an important step would
be to construct a model for such a matter and see what mechanism
could be at work. Clearly, the model must be manifestly confining,
chirally symmetric and provide dynamical breaking of chiral 
symmetry. Constructions that are based on the NJL model or the
linear sigma model and their extensions are not suited because
of luck of  confinement of quarks. 

The
simplest possible model that satisfies all required criteria is the model
of refs. \cite{Y,AD}. This 3+1 dimensional model can be considered 
as a straightforward
generalization of the QCD in 1+1 dimension in the large $N_c$ limit
- the 't Hooft model \cite{H}. It is assumed within the model that
the only gluonic interaction between quarks is an instantaneous linear
potential of Coulomb type.
Such a potential is indeed observed in variational \cite{SW} as well
as lattice Coulomb gauge simulations \cite{V}.
 An important aspect of this model is that
it exhibits effective chiral restoration in hadrons with large spins J
 \cite{largeJ,G4}.\footnote{The chiral symmetry 
 breaking is important only for quarks
with low momenta.
 At large J the low
momenta of quarks in hadrons are cut off by the centrifugal repulsion,
however.
Consequently, the chiral symmetry breaking in the
vacuum is irrelevant to the mass generation of hadrons with large $J$.}

This model was used in ref. \cite{GW} to answer the
question about possible mechanism that could be responsible for the confing
but chirally symmetric dense, cold  matter. Assuming a liquid phase,
i.e., that the rotational and translational symmetries are not broken
(as it is in the nuclear matter), the Pauli blocking of the positive energy 
levels by valence quarks prevents
the gap equation to generate a nontrivial solution with broken chiral
symmetry above some critical quark Fermi momentum. At the same time 
confinement still persists in the system, so
that the uncorrelated colored single quark excitations are
not possible.

In the latter work a rigid quark Fermi surface was assumed, like for the
noninteracting fermions at $T=0$. If such a phase exists, however,
relevant degrees of freedom near the Fermi surface are baryons.
Quarks interact inside baryons. Consequently, the quark distribution
function near the Fermi surface must be smooth. Here we study effects of such
 diffusion of the quark Fermi surface and solve the corresponding
gap and Bethe-Salpeter equations. We conclude that for any reasonable
diffusion there always exists such a critical "Fermi momentum" of quarks at
which the chiral restoration phase transition persists and, hence, 
elementary excitations above this critical "Fermi momentum" are
the color singlet chirally symmetric hadron modes.

\section{Overview of confinement and chiral symmetry breaking in a vacuum}

The $SU(2)_L \times SU(2)_R  \times U(1)_A \times U(1)_V $ symmetric Hamiltonian

\begin{eqnarray} 
\hat{H} & = & \int d^3x\bar{\psi}(\vec{x},t)\left(-i\vec{\gamma}\cdot
\vec{\bigtriangledown} \right)\psi(\vec{x},t) \nonumber \\
 &+& \frac12\int d^3
xd^3y\;J^a_\mu(\vec{x},t)K^{ab}_{\mu\nu}(\vec{x}-\vec{y})J^b_\nu(\vec{y},t),
\label{H} 
\end{eqnarray}

\noindent
 relies on the quark current-current 
 $J_{\mu}^a(\vec{x},t)=\bar{\psi}(\vec{x},t)\gamma_\mu\frac{\lambda^a}{2}
\psi(\vec{x},t)$ interaction
via the instantaneous linear interquark potential of the Coulomb
type

\begin{equation} 
K^{ab}_{\mu\nu}(\vec{x}-\vec{y})=g_{\mu 0}g_{\nu 0}
\delta^{ab} V (|\vec{x}-\vec{y}|); ~~~~~
\frac{\lambda^a \lambda^a}{4}V(r) = \sigma r,
\label{KK}
\end{equation}

\noindent
where $a,b$ are color indices.
This model was intensively used in the past to study chiral symmetry
breaking, chiral properties of hadrons, etc, see, for example, 
refs. \cite{Y,AD,bic}.

 The Fourier transform of the linear
potential and any loop integral are not defined in the infrared
region, $ p \sim 0$.
Hence,  an infrared regularization is required.
Physical color singlet observables, such as  hadron masses, 
chiral condensate, etc., must be independent of the infrared
regulator $\mu_{IR}$ in the infrared limit  $\mu_{IR} \rightarrow 0$.

There are several physically equivalent ways to perform this infrared
regularization. Here we follow Ref. \cite{Alkofer} and define the
potential in momentum space as

\begin{equation}
V(p)= \frac{8\pi\sigma}{(p^2 + \mu_{\rm IR}^2)^2}.
\label{FV} 
\end{equation}

\noindent
This potential in the configuration space contains the
required $\sigma r$ term,  the infrared divergent term
$-\sigma /\mu_{IR}$ as well as terms that vanish in the infrared limit.

The self-energy operator 

\begin{equation}
\Sigma(\vec p) =A_p+(\vec{\gamma}\hat{\vec{p}})(B_p-p)
\label{SE} 
\end{equation}

\noindent
consists of the Lorentz-scalar chiral symmetry breaking part
$A_p$ and chirally symmetric part $(\vec{\gamma}\hat{\vec{p}})(B_p-p)$.
The unknown functions $A_p$ and $B_p$ are to be determined from
the gap equation for the chiral angle
 $\varphi_p$
 
 \begin{equation}
 A_p \cos \varphi_p - B_p \sin \varphi_p = 0,
 \label{gap}
 \end{equation}

\noindent
where
 
\begin{equation}
A_p  = \frac{1}{2}\int\frac{d^3k}{(2\pi)^3}V
(\vec{p}-\vec{k})\sin\vp_k,
\label{Ap}
 \end{equation}

\begin{equation}
B_p  =  p+\frac{1}{2}\int \frac{d^3k}{(2\pi)^3}\;(\hat{\vec{p}}
\hat{\vec{k}})V(\vec{p}-\vec{k})\cos\vp_k. 
\label{Bp} 
\end{equation} 

These integrals contain both the infrared divergent and the
infrared finite parts

 \begin{equation}
A_p=\frac{\sigma}{2\mu_{\rm IR}}\sin\varphi_p+A^f_p, 
\label{AA}
\end{equation}
 
 \begin{equation}
B_p=\frac{\sigma}{2\mu_{\rm IR}}\cos\varphi_p+B^f_p.
\label{BB}
\end{equation}

\noindent
The same is true for the single quark energy:

 \begin{equation}
\omega_p = \sqrt(A_p^2 +B_p^2) = 
\frac{\sigma}{2\mu_{\rm IR}} +\omega^f_p. 
\label{AA}
\end{equation}

\noindent
Consequently, the single quark Green function is divergent
and the single quark energy is infinite in the infrared
limit. Actually, energy of any color nonsinglet state is infinite.
 At the same time the infrared divergence exactly cancels
out in any color singlet quantity and these quantities are
finite and well defined \cite{GW}.
This is a manifestation of confinement within this
model.

Similarly, the infrared divergence in $A_p$ and $B_p$ cancels
in the gap equation and this equation can be solved directly
in the infrared limit. The gap equation can be solved numerically
and a nontrivial solution with the chiral (Bogoliubov) angle
$\varphi_p \neq 0$ signals 
dynamical breaking of chiral symmetry.  Hence, the single quark
Green function is not chirally symmetric, $A_p \neq 0$; there
appears the quark condensate and dynamical mass of quarks

\begin{equation}
\langle\bar{q}q\rangle=-\frac{N_C}{\pi^2}\int^{\infty}_0 dp\;p^2\sin\vp_p,
~~~~~~~
M(p) = p \tan \varphi_p.
\label{dyna}
\end{equation} 

\section{Chiral symmetry breaking and confinement in a dense 
matter at T=0 with a rigid quark Fermi surface}

It is practically impossible to solve exactly the model in a dense
matter. Indeed, that would imply to solve it first for a single baryon; then
to obtain a baryon-baryon interaction; given this interaction to construct
a nuclear matter and then slowly to increase its density. Obviously,
it is a formidable problem. In order to proceed and get some
insight one needs justifiable simplifications.

In the large $N_c$ limit the nucleon is infinitely heavy, translational
invariance is broken and a many-nucleon system is certainly in a crystal
phase. Whether a (dense) nuclear matter will be a liquid or a crystal
at $N_c=3$ is a subject to dynamical calculations. Such
microscopical calculations cannot be persued for any "realistic" model in 3+1
dimensions with
confinement and (broken) chiral symmetry. However, in the real world 
$N_c=3$ we do know that the nuclear
matter is in a liquid phase; both translational and rotational
invariances are intact. We then assume a liquid phase with manifest
translational and rotational invariances in a dense quarkyonic matter. 

We also assume that confinement persists up to  large densities
 at $N_c=3$.  At $T=0$ deconfinement could
happen through the Debye screening of the confining gluon propagator:
A gluon creates the quark - quark hole pair that again annihilates into
a gluon. If this vacuum polarization diagram is finite, then at some
density there should happen a complete screening of the confining
gluon field. However, in the confining mode  energy of the
colored quark - quark hole pair is infinite  \cite{guo}. 
The allowed excitations
in the confining mode are the color singlet excitations like baryon - baryon
hole pairs, etc. These excitation modes cannot screen the confining colored
gluon propagator. In this sense the $T=0$ physics is rather different
from the deconfinement at zero density and large temperature. In the
latter case a screening proceeds via the incoherent thermal gluon 
loops.

One could expect that the deconfinement
should happen in a dense medium at $T=0$ due to perlocation of baryons.
 Such a reasoning is too
naive, however, because the perlocation does not yet imply screening
of the confining gluon field. For example, deconfinement never happens
at $T=0$ in QCD 
at large $N_c$ or in the
't Hooft model. In both cases  baryons "sit"
on top of each other in a very dense medium, but it is still a system
with confinement.

We want to address the chiral symmetry breaking properties of
a dense matter with confinement. In the vacuum dynamical chiral
symmetry breaking happens because there is an attractive interaction
between the left quarks and the right antiquarks and vice versa. 
This attractive
interaction shifts  the energy of the vacuum with broken
chiral symmetry  below the energy of the perturbative Dirac vacuum. 

Consequently, in a dense matter at $T=0$ the most important
physics that  leads to the restoration of chiral symmetry
is the Pauli blocking  (by  the valence quarks) of the positive energy 
levels required for the very existence of the
quark condensate.

\begin{figure}[h]
  \begin{center}
    \begin{minipage}[h]{0.4\linewidth}
      \center{\includegraphics[width=1\linewidth]{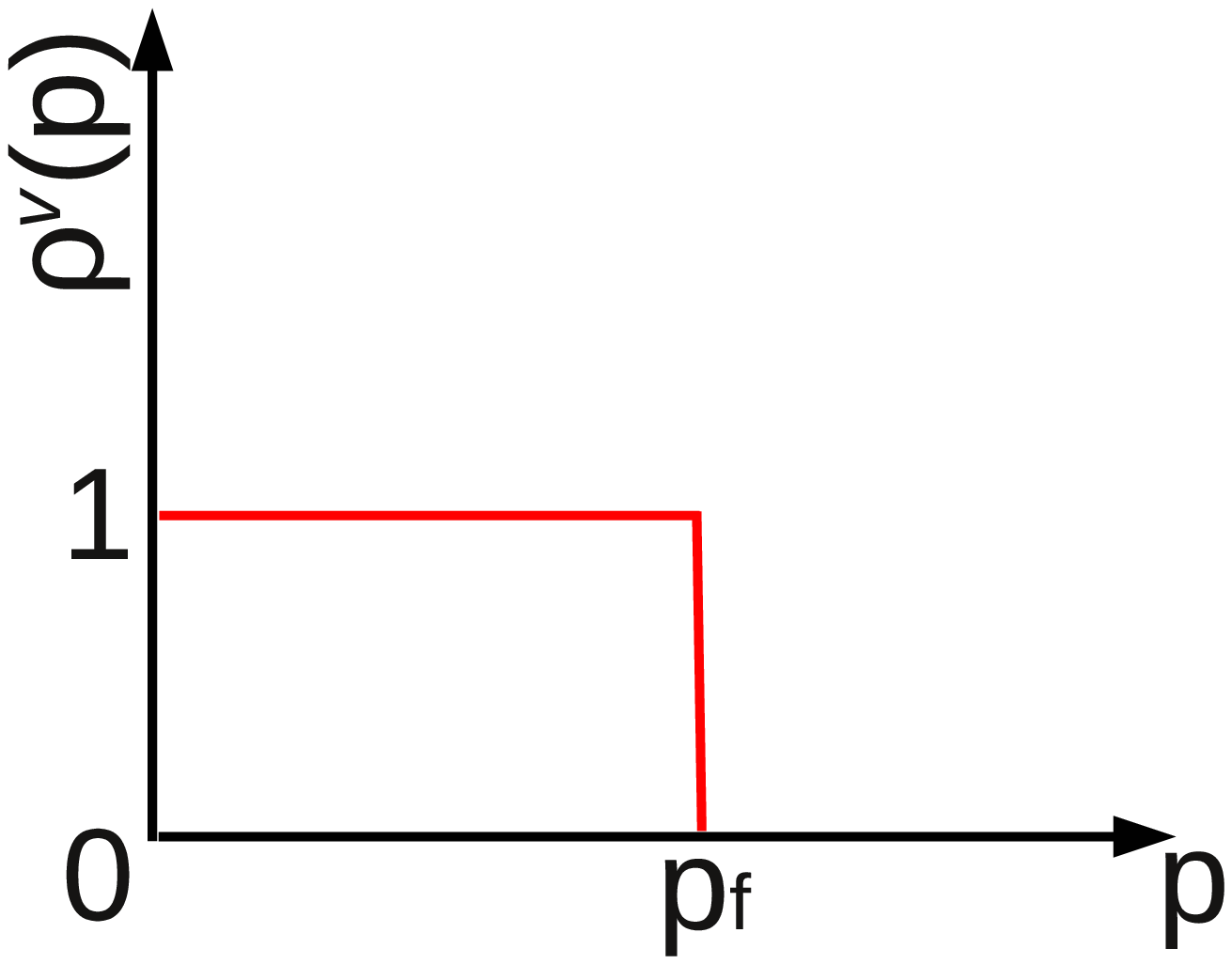} \\ a)}
    \end{minipage}
    \hspace*{20pt}
    \begin{minipage}[h]{0.4\linewidth}
      \center{\includegraphics[width=1\linewidth]{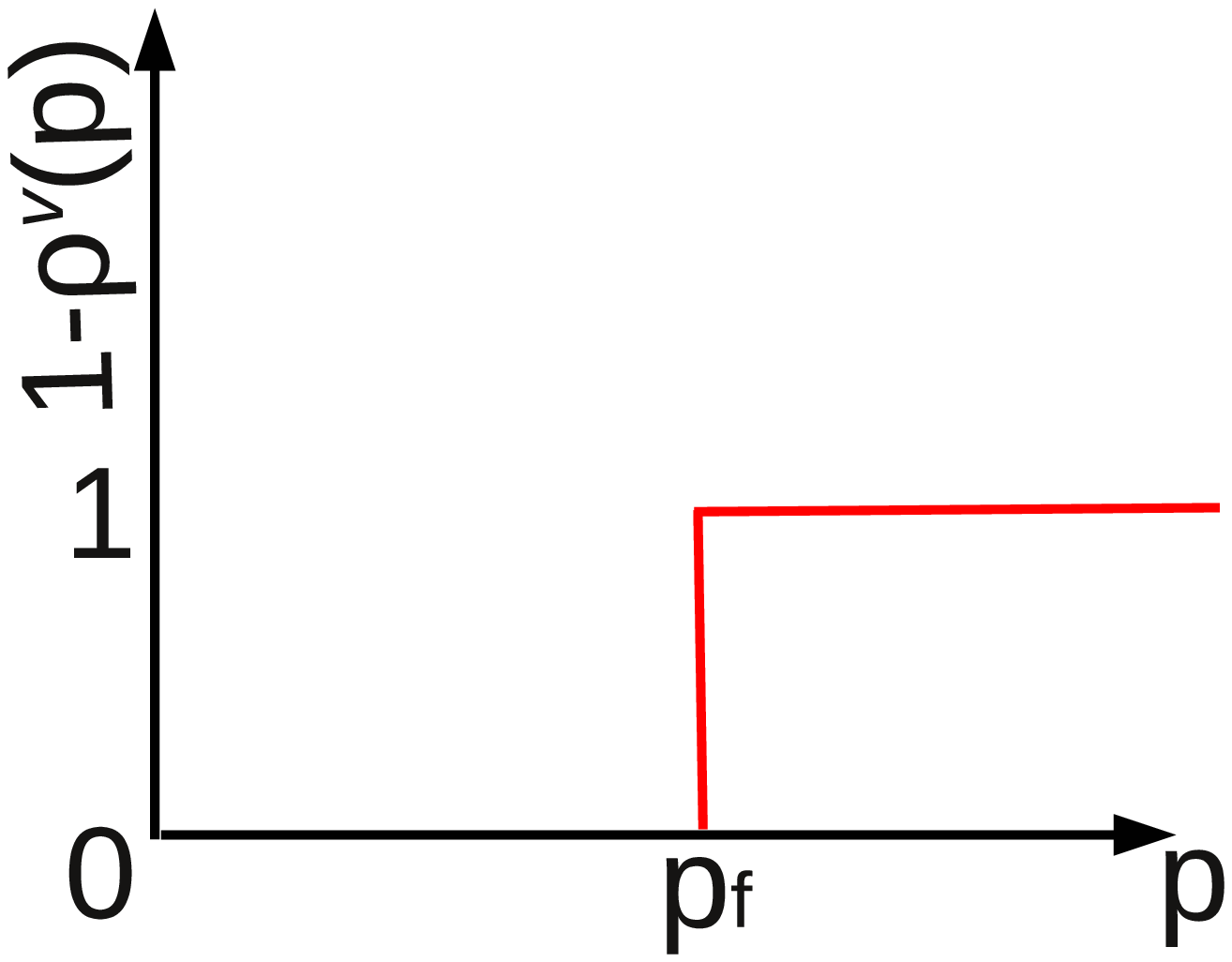} \\ b)}
    \end{minipage}
  \end{center}
  \caption{Valence quark distribution and 
  the corresponding integration weight.}
\end{figure}

In our previous work \cite{GW} the effect of the Pauli blocking
was studied assuming a rigid valence quark Fermi sphere, like
for the ideal Fermi gas at $T=0$, see Fig. 1a. In this case all  quark 
positive energy levels
below the Fermi momentum $p_f$ are occupied by the valence quarks
and
one has to replace the vacuum density matrix $v(\vec p) v^\dagger(\vec p)$
by the density matrix in the medium:

\begin{equation}
\rho(\vec p) = \Theta(p_f - p) u(\vec p) u^\dagger(\vec p)
+ v(\vec p) v^\dagger(\vec p).
\label{den}
\end{equation}

\noindent
Hence, within the mean field approximation
one has to remove from the integration
in the gap equation all
quark momenta below $p_f$ since they are Pauli  blocked, see Fig. 1b. 
The modified gap equation  
is then the same as in (\ref{gap}) - (\ref{Bp}),
but the integration starts not from $k=0$, but from $k=p_f$.

 \begin{figure}
 \includegraphics[width=0.88\hsize,clip=]{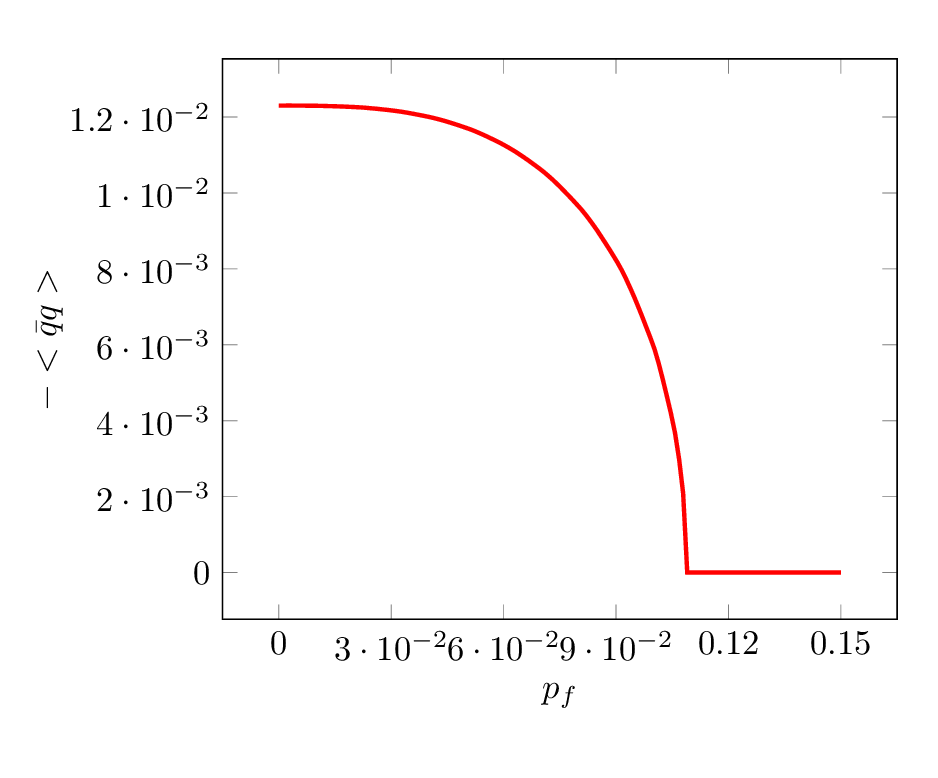}
 \caption{Quark condensate in units of $\sigma^{3/2}$
 as a function of the Fermi momentum, which is units of $\sqrt \sigma$.}
 \end{figure}

At the critical Fermi momentum $p_f^{cr} \sim 0.109 \sqrt\sigma$ 
the chiral phase transition
is observed, see Fig. 2, because the nontrivial solution with broken chiral
symmetry disappears. Above this phase transition the chiral angle $\varphi_p$,
the quark condensate $\langle \bar q q \rangle$, the dynamical mass
of quarks $M(p)$ as well as the chiral symmetry breaking part $A_p$ of the
quark Green function identically vanish. At the same time the chirally
symmetric part $B_p$ of the quark Green function does not vanish and is
still in fact infrared divergent. The single quark energy is 
infinite and confinement persists even in the chirally symmetric phase.
The color singlet hadronic excitations have finite and well defined
energy.

\section{Effects of a diffusion of the quark Fermi surface}

\begin{figure}[h]
  \begin{center}
    \begin{minipage}[h]{0.4\linewidth}
      \center{\includegraphics[width=1\linewidth]{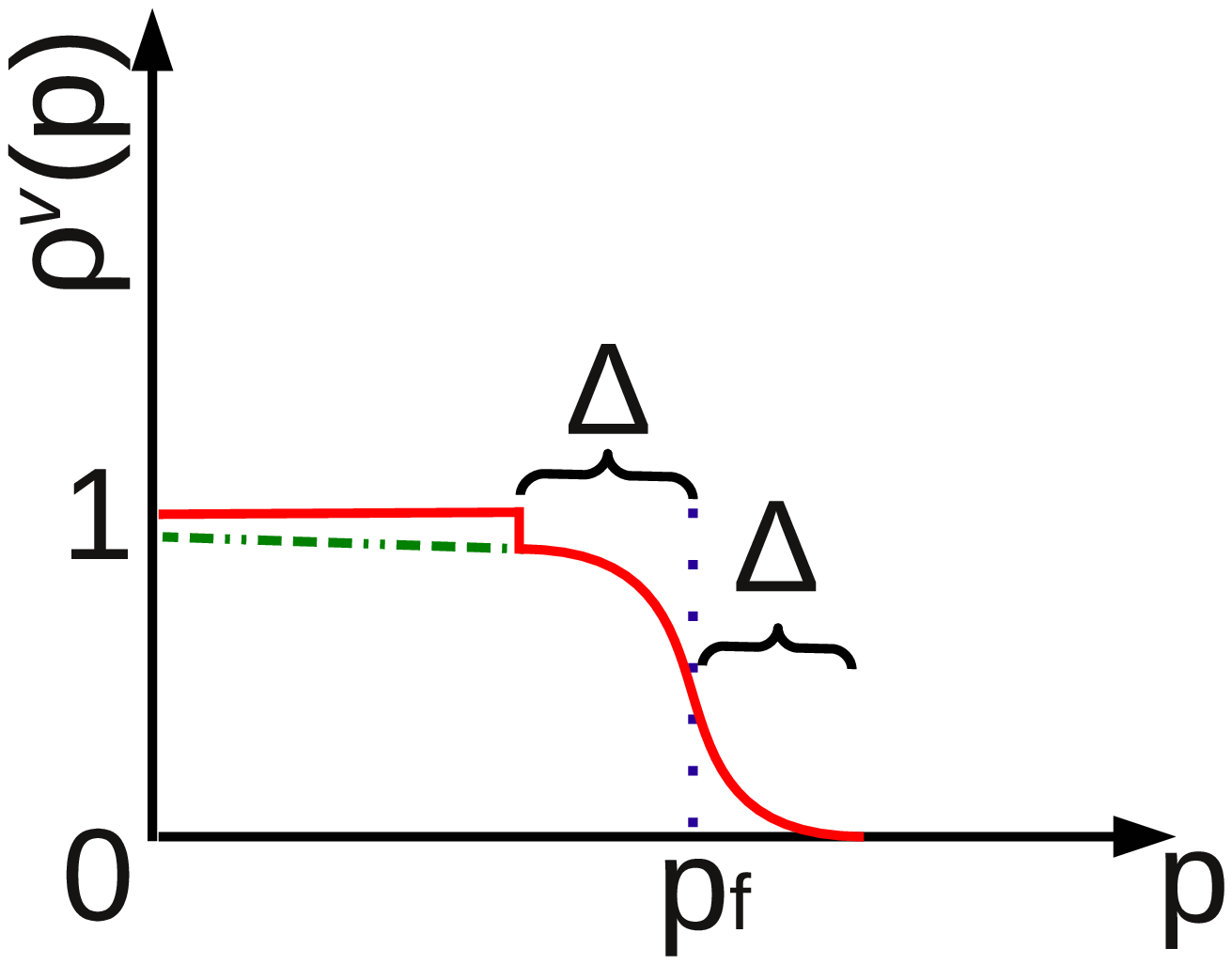} \\ a)}
    \end{minipage}
    \hspace*{20pt}
    \begin{minipage}[h]{0.4\linewidth}
      \center{\includegraphics[width=1\linewidth]{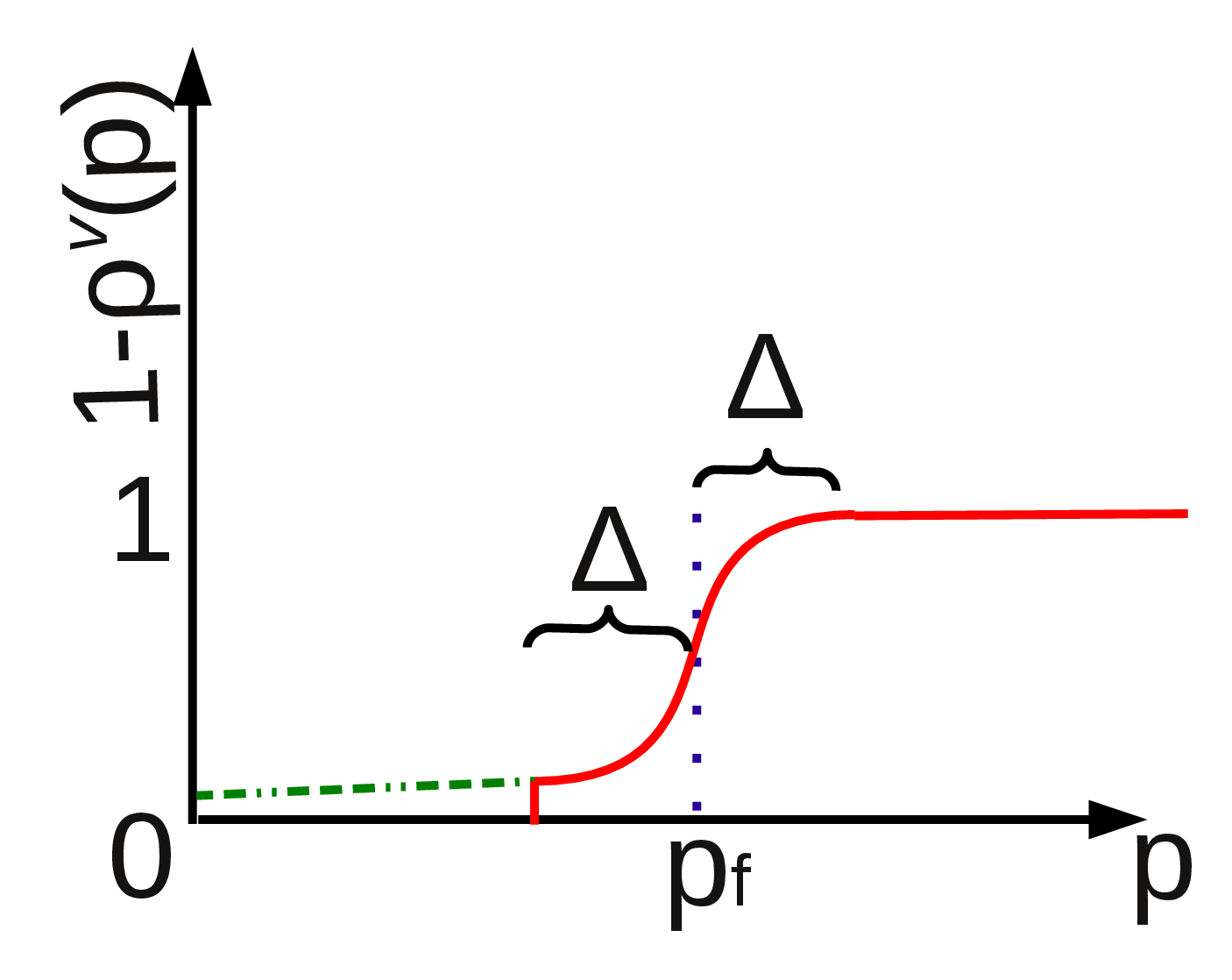} \\ b)}
    \end{minipage}
  \end{center}
  \caption{Diffused step valence quark distribution and the
  corresponding integration weight.}
\end{figure}

In reality valence quarks near the Fermi surface interact and cluster into 
the color 
singlet baryons. This interaction in general would lead
to a diffusion of the rigid
Fermi surface for quarks. Some levels above the "Fermi momentum" must
be occupied with some probability as well as some levels below the
"Fermi momentum" with some probability must be empty. 

In principle, the quark distribution function near the diffused Fermi
surface could be obtained selfconsistently from the full solution of 
the problem.
It is a formidable task and such a program cannot be persued.
However, it is  clear that the realistic distribution function
will be smooth, of the form on Fig. 3a. Then, for our present goal it will
 suffice to parameterize such a smooth distribution in a simple form
and study effect of the diffusion on the solution of the gap equation.

We parameterize a smooth valence quark distribution function by
\begin{equation}
  \rho^v(p) = \Theta(-p + p_f - \Delta) +
  \Theta(p - p_f + \Delta)\frac{1}{e^{(p - p_{f}) / \Delta} + 1}. 
  \label{rho}
\end{equation}

\noindent
Given the valence distribution function  we multiply
the integrands in eqs. (\ref{Ap}) - (\ref{Bp}) by the weight
function $1 - \rho^v(k)$ and solve the gap equation for
different $p_f$ and diffusion width $\Delta$.

If
the diffusion width is much smaller than the critical
Fermi momentum for a rigid Fermi surface, $\Delta \ll p_f^{cr}(\Delta=0)$,
(what should be considered as a realistic situation),
then the evolution of the chiral condensate with $p_f$ is
similar to the case of the rigid quark Fermi surface.
This situation is represented by the curve $\Delta = 0.02$
on Fig. 4. The phase transition happens at the "Fermi momenta" that
are rather close to the critical Fermi momentum, $p_f^{cr}(\Delta=0) = 0.109$
from Fig. 2 (see also the curve $\Delta = 0$ on Fig. 4). This
can be easily understood. At all  momenta  $p \ll p_f$ the Pauli
blocking on Fig. 3 is the same as for the rigid quark Fermi surface.
At momenta just below the $p_f$ the effect of the Pauli blocking
is weaker than for the rigid Fermi surface. However, this is 
 compensated by additional Pauli blocking of the levels that
are just above the Fermi momentum for the rigid quark distribution.
Consequently, with a small diffusion widths the "critical Fermi momentum",
$p_f^{cr}(\Delta)$,
at which the phase transition happens, is shifted to slightly  lower
values of $p_f$.

However, if a diffusion width becomes larger and eventually comparable
with the "Fermi momentum", then the "critical Fermi momentum",
at which the phase transition happens, increases.

For each fixed diffusion width $\Delta$ there always exist such
 "Fermi momenta" where the Wigner-Weyl mode of chiral symmetry is
 realized. This can be seen from Fig. 5, where a line of "critical
 Fermi momenta" is depicted. The area above this critical line
 corresponds to the chirally symmetric phase, while all points
 below the critical line represent a matter with broken chiral
 symmetry.

\begin{figure}
\includegraphics[width=0.88\hsize,clip=]{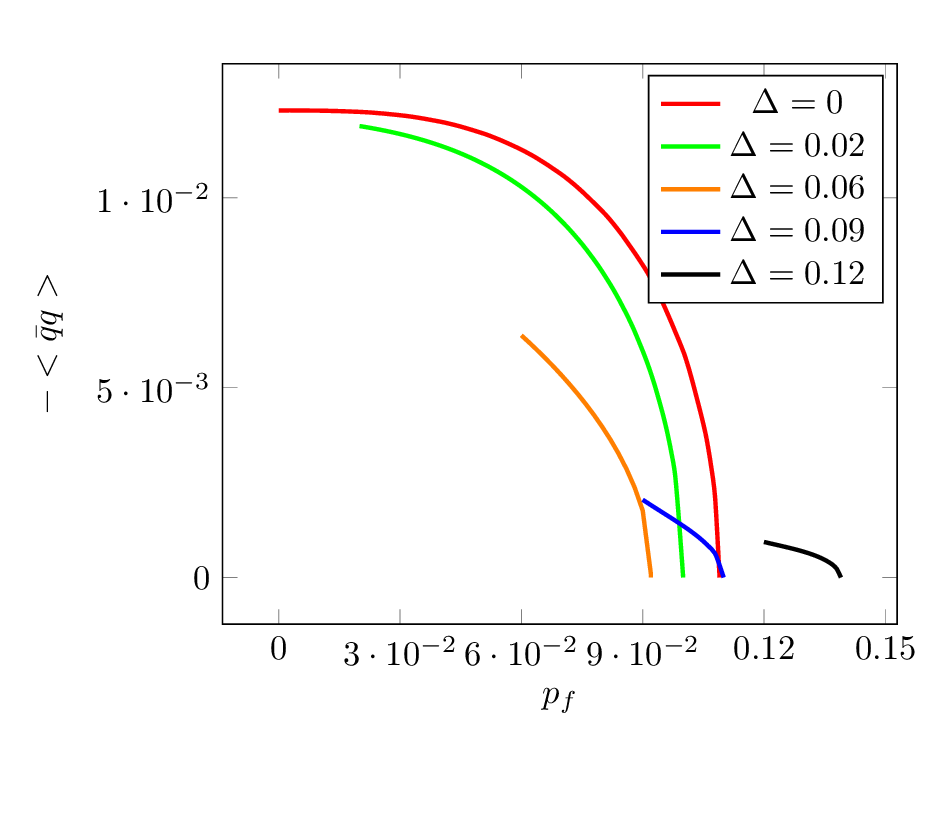}
\caption{Quark condensate (in units of $\sigma^{3/2}$)
as a function of the Fermi momentum (in units of $\sqrt \sigma$)
and the diffusion width $\Delta$ (in units of $\sqrt \sigma$) }
\end{figure}

\begin{figure}
\includegraphics[width=0.88\hsize,clip=]{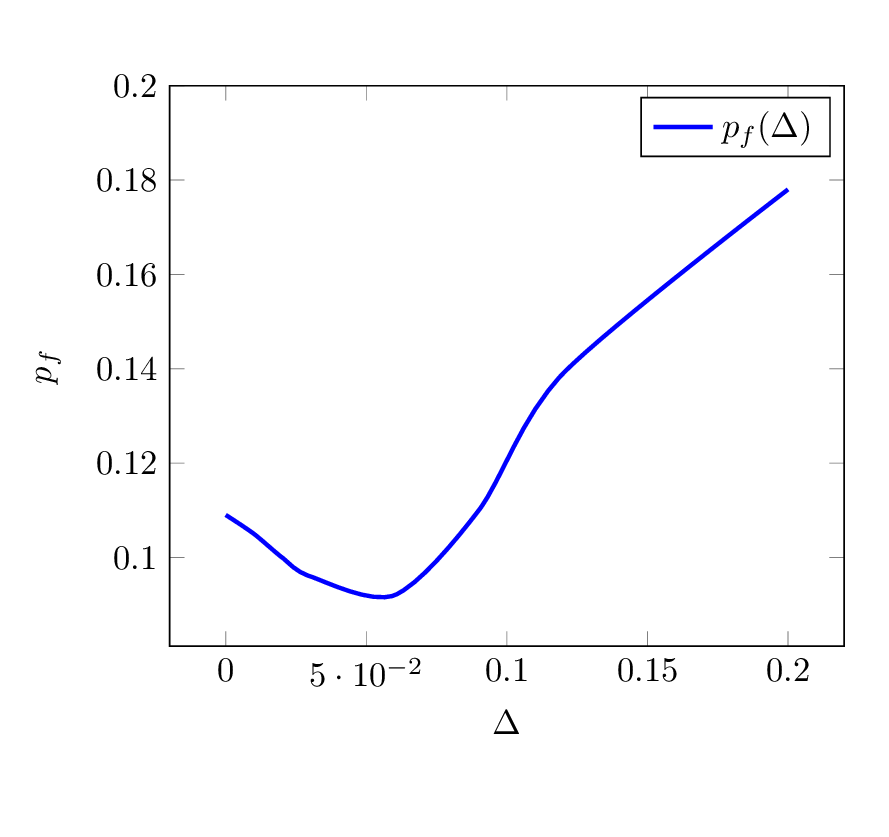}
\caption{Critical line that separates the quarkyonic matter
with broken and restored chiral symmetry.}
\end{figure}

The chiral angle $\varphi_p$ and dynamical mass of quarks $M(p)$ in
the Nambu - Goldstone mode of chiral symmetry  at some
"Fermi momentum" and different diffusion widths $\Delta$  
 are shown on Figs. 6-7.

\begin{figure}
\includegraphics[width=0.88\hsize,clip=]{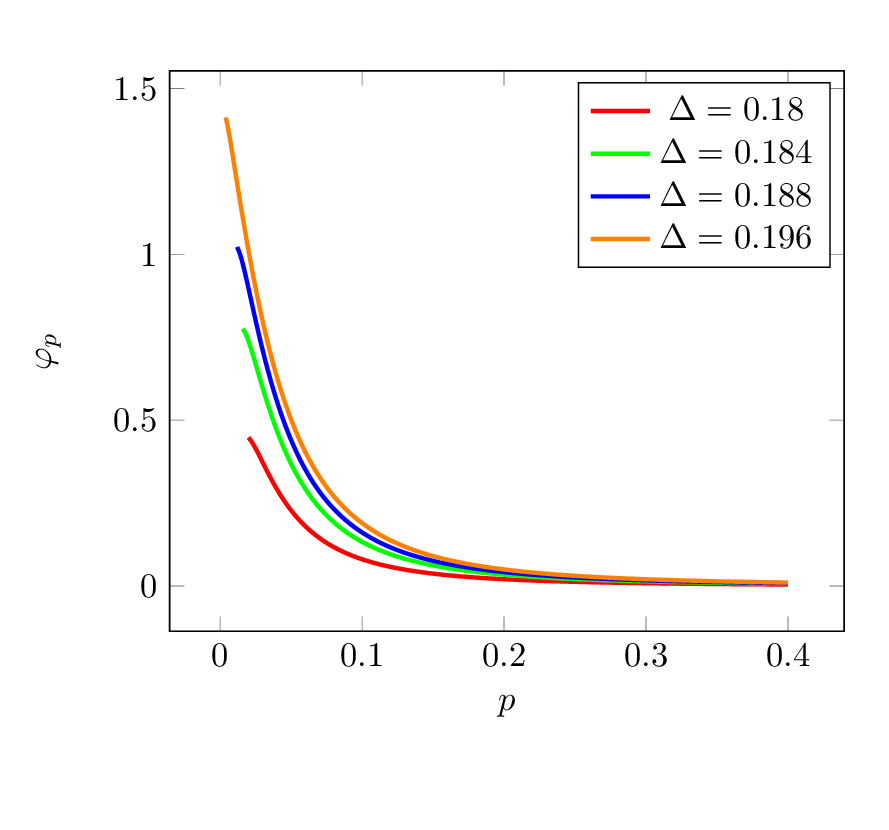}
\caption{Chiral angle $\varphi_p$ as a function of the momentum $p$
for a "Fermi momentum" $p_f=0.2$ at different fixed values of
smoothing $\Delta$. $p$ and  $\Delta$ are in units of $\sqrt \sigma$. }
\end{figure}

\begin{figure}
\includegraphics[width=0.88\hsize,clip=]{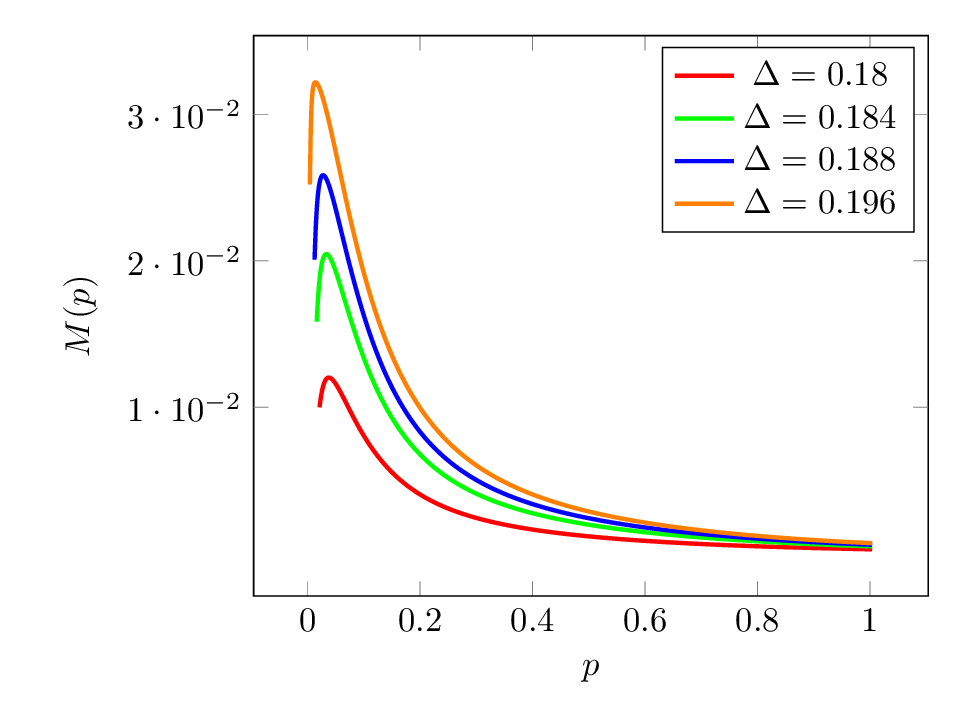}
\caption{Dynamical mass $M(p)$ as a function of the momentum $p$
for a "Fermi momentum" $p_f=0.2$ at different fixed values of
smoothing $\Delta$. $M(p)$, $p$ and  $\Delta$ are in units of $\sqrt \sigma$. }
\end{figure}

\section{Chiral restoration in meson spectra}

Another explicit illustration of chiral symmetry 
of a dense matter  above the chiral restoration
phase transition are  properties of hadronic
excitations. In the Nambu-Goldstone mode of chiral symmetry
there must be a massless
excitation mode that is associated with the massless pion. At the
same time energies of all other  mesons must be finite.
In particular, there must be a finite splitting of the excitations
with quantum numbers $I, J^{PC} = 1, 0^{-+}$  and $I, J^{PC} = 0, 0^{++}$,
that will be reffered as the pion and the $\sigma$-meson, respectively, according
to the standard nomenclature. In contrast, these excitations must be
exactly degenerate in the Wigner-Weyl mode of chiral symmetry and
form the $(1/2,1/2)_a$ representation of the $SU(2)_L \times SU(2)_R$
chiral group \cite{G4}.

To obtain the quark-antiquark bound states we solve the homogeneous
Bethe-Salpeter equation in the rest frame

\begin{eqnarray}
\chi(m,\vpp)&= &- i\int\frac{d^4q}{(2\pi)^4}
V(|\vpp-\vq|)\;
\gamma_0 S(q_0+m/2,\vpp-\vq) \nonumber \\
& \times & \chi(m,\vq)S(q_0-m/2,\vpp-\vq)\gamma_0(1 -\rho^v(q)).
\label{GenericSal}
\end{eqnarray}

\noindent
Here $S$ is the dressed single quark propagator, that is the solution
of the gap equation,
 $m$ is the meson mass and $\vec p$ is the relative momentum. 
The Bethe-Salpeter
equation is solved by means of expansion of the vertex function
$\chi(m,\vpp)$  into a set of all possible independent 
amplitudes consistent with $I,J^{PC}$ and it
transforms into a system of coupled equations. The infrared
divergence cancels exactly in these equations and they can
be solved numerically \cite{largeJ}. The Pauli blocking by valence
quarks is taken into account via the weight function $1 - \rho^v(q)$.

\begin{figure}
\includegraphics[width=0.6\hsize,clip=]{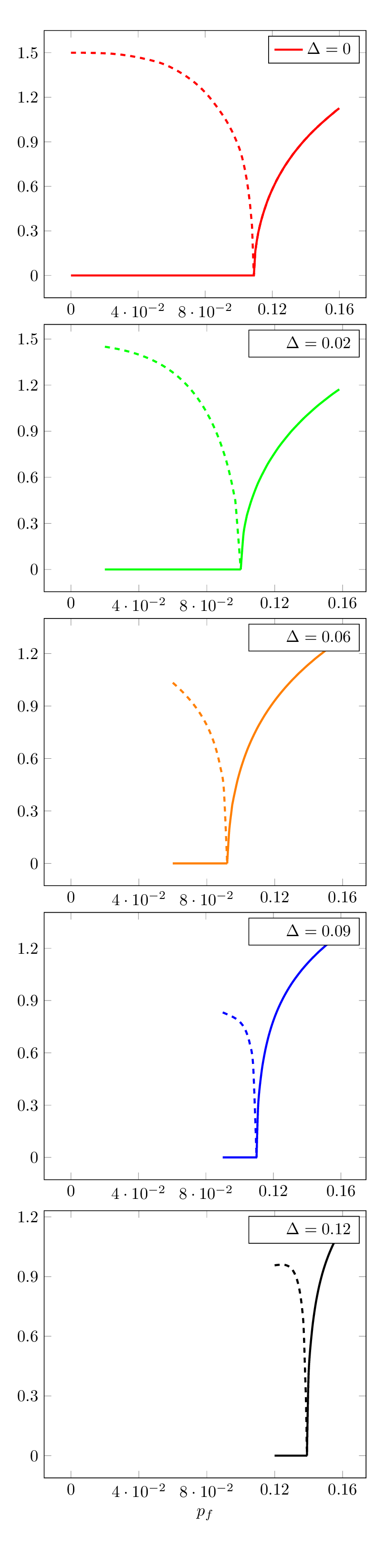}
\caption{Masses of the pseudoscalar (solid) and scalar (dashed) mesons 
in units of $\sqrt \sigma$ as functions of the "Fermi momentum" $p_f$
and of the diffusion width $\Delta$ (in units of $\sqrt{\sigma}$).}
\end{figure}

In the Wigner-Weyl mode, {\it i.e.}, when 
dynamical quark mass and chiral angle vanish, $M(p) = 0; \varphi_p = 0$, 
the Bethe-Salpeter equations
for the $1, 0^{-+}$ and $0, 0^{++}$ bound states
become identical \cite{largeJ} and consequently energies of these
states coincide.

On Fig. 8 we show masses of both pseudoscalar and scalar modes
for different "Fermi momenta" $p_f$ and diffusion widths $\Delta$.
For each $\Delta$ there is a critical $p_f^{cr}(\Delta)$ at which the
chiral restoration phase transition takes place. Below this 
 $p_f^{cr}(\Delta)$ there is a massless pion and a massive $\sigma$-meson. 
 Above
the critical $p_f$ both the pion and the $\sigma$-meson are massive
and exactly degenerate.

\section{Conclusions}

In the confining mode the valence quarks interact and near the Fermi surface 
cluster into the color singlet baryons. This implies that there cannot
be a rigid quark Fermi surface. The valence quark distribution
function near the Fermi surface must be smooth. The valence quark
levels above the "Fermi momentum" are occupied with some probability
as well as the levels below the "Fermi momentum" must be, with some 
probability, empty. We assume unbroken translational and
rotational invariances, {\it i. e.}, a liquid phase. We
parameterize such a diffused "Fermi surface" by a
simplest possible function and solve the corresponding gap and
Bethe-Salpeter equations. By this we  verify whether a chiral phase transition,
previously observed for a rigid quark Fermi surface, survives or not.
It turns out that for any reasonable diffusion width there always exists
such a "Fermi momentum" that the chiral restoration phase transition
does take place. This reconfirms our previous conclusions about
possible existence of the confining but chirally symmetric phase.
Below the phase transition the elementary excitation modes of a matter
are hadrons with broken chiral symmetry, while above the phase transition
such excitations are chirally symmetric hadrons.

\medskip
{\bf Acknowledgments}
Support of the Austrian Science
Fund through the grant P21970-N16 is acknowledged.

\end{document}